\documentclass{article}
\usepackage{spconf,amsmath}
\usepackage{amsthm}
\usepackage{amsfonts}
\usepackage{amssymb}
\usepackage{mathrsfs}
\usepackage{mathtools}
\usepackage[english]{babel}
\usepackage{float}
\usepackage[toc,page]{appendix} 
\usepackage[pdftex]{graphicx}
\usepackage{epstopdf}
\usepackage{amsmath}
\usepackage{color}
\usepackage{multirow}
\usepackage{graphicx}
\usepackage{IEEEtrantools}
\usepackage{bm}
\usepackage{balance}
\usepackage{microtype}
\usepackage[caption=false,font=footnotesize]{subfig}
\usepackage{cite}
\usepackage{comment}

\usepackage[switch]{lineno}
\usepackage[named]{algo}
\usepackage[noend]{algpseudocode}
\usepackage{algorithm}

\ninept

\newcommand{\dddag}{%
  \mathbin{\vbox{\offinterlineskip\ialign{%
    \hfil##\hfil\cr
    \small{$\dagger$}\cr
    \noalign{\kern-0.6ex}
    \small{$\ddagger$}\cr
}}}}

\title{OptM3Sec: Optimizing Multicast IRS-Aided Multiantenna DFRC Secrecy Channel with Multiple Eavesdroppers} 
	
\name{Kumar Vijay Mishra$^{\dag}$, Arpan Chattopadhyay$^{\ddag}$, Siddharth Sankar Acharjee$^{\ddag}$ and Athina P. Petropulu$^{\dddag}$}
\address{$^{\dag}$United States CCDC Army Research Laboratory, Adelphi, MD 20783 USA\\ 
$^{\ddag}$Indian Institute of Technology, New Delhi 110016 India\vspace{-6pt}\\ 
$^{\dddag}$Rutgers - The State University of New Jersey, Piscataway, NJ 08854 USA\vspace{-12pt}}

\begin{document}
\setlength{\abovedisplayskip}{3pt}
\setlength{\belowdisplayskip}{3pt}

\maketitle
\vspace{-12pt}
\begin{abstract}
With the use of common signaling methods for dual-function radar-communications (DFRC) systems, the susceptibility of eavesdropping on messages aimed at legitimate users has worsened. For DFRC systems, the radar target may act as an eavesdropper (ED) that receives a high-energy signal thereby leading to additional challenges. Unlike prior works, we consider a multicast multi-antenna DFRC system with multiple EDs. We then propose a physical layer design approach to maximize the secrecy rate by installing intelligent reflecting surfaces in the radar channels. Our \textit{opt}imization of \textit{m}ultiple ED \textit{m}ulticast \textit{m}ulti-antenna DFRC \textit{sec}recy rate (OptM3Sec) approach solves this highly nonconvex problem with respect to the precoding matrices. Our numerical experiments demonstrate the feasibility of our algorithm in maximizing the secrecy rate in this DFRC setup.
\end{abstract}

\vspace{-12pt}
\section{Introduction}
With the advent of mobile communications, portions of spectrum earlier solely assigned to radar are being re-allocated for dual-use with communications 
\cite{griffiths2015radar}. This has accelerated efforts to allow uncontested joint access to the bandwidth through coexisting radar and communications systems, apart from other emitters within the same RF spectrum \cite{mishra2019toward}. In this context, dual-function radar-communications (DFRC) systems have emerged as a viable and relatively less contentious approach because it hosts both transmissions on the same hardware unit using same frequency band \cite{hassanien2015dual}. Since DFRC employs an identical waveform for both services, the delivery of information to both communications users and radar targets has increased chances of unauthorized users or \textit{eavesdroppers} (EDs) intercepting it \cite{chalise2018performance}. In this paper, we focus on secure DFRC transmission.

In general, cryptographic techniques are effective for secure transmission  \cite{schurmann2011secure}. However, with the computational  power available today, it is possible to break cryptographic codes \cite{hamamreh2018classifications}. A complementary approach lies in ensuring physical layer security so that EDs are not able to decode the message successfully, even if they knew the secret code. The physical layer secrecy seeks to maximize, by exploiting channel conditions,  the rate of reliable information delivery to the intended receiver, with the ED being kept as ignorant of that information as possible. This line of research was pioneered by Wyner \cite{Wyner75}, who introduced the wiretap channel and the notion of {\it secrecy capacity}, i.e., the rate at which the legitimate receiver correctly decodes the source message, while an ED obtains no useful information about the source signal. For the classical source-destination-ED Gaussian wiretap channel, the secrecy capacity is zero when the quality of legitimate channel is worse than eavesdropping channel \cite{Leung78}. One way to achieve non-zero secrecy rates is to adopt \textit{cooperative jamming} by introducing
one \cite{Fakoorian2011solutions,Lingxiang14,Gan13,zheng2015secrecy} or more
\cite{Zheng2011optimal,li2011oncooperative,dong2010improving,luo2013uncoordinated,Hoon2014multiuser} relays (helpers), which transmit artificial noise in a beamforming fashion thereby acting as jammers for the purpose of degrading the channel to the ED. Alternatively, the source may embed artificial noise in its transmission and beamformed it not to interfere with the legitimate receiver \cite{khisti2010secure,goel2008guaranteeing}.

In a DFRC system, 
the radar target is illuminated by a high-energy signal, which also contains information intended for the communication users. 
This increases the susceptibility of eavesdropping on the communications messages and may be aggravated when DFRC employs multiple-input multiple-output (MIMO) antennas \cite{su2020secure} because of its omnidirectional radiation pattern. Recent studies show that physical layer security could be a promising solution for a secure MIMO DFRC  \cite{eltayeb2017enhancing} by trading-off the secrecy rate (the achievable difference between the communication rates to the legitimate users and the target) for the data rate. The MIMO communications-radar system presented in \cite{Deligiannis_Secrecy_Radar_TAES18} simultaneously transmitted jamming and communications messages to confuse an ED (co-located with the targets) while trying to enhance target detection. In \cite{su2020secure}, the MIMO DFRC employed artificial noise and minimized the signal-to-interference-and-noise ratio (SINR) at the eavesdropping target with an SINR constraint at the legitimate users. These systems were extended in \cite{fangsinr} by incorporating an intelligent reflecting surface (IRS) in the channel. An IRS is a two-dimensional surface consisting of a large number of passive meta-material elements to reflect the incoming signal through a pre-computed phase shift \cite{elbir2020survey}. The IRSs have emerged as feasible low cost, light-weight, and compact alternatives to large arrays for both communications \cite{hodge2020intelligent,ahmed2022joint} and radar \cite{esmaeilbeig2021irs}.

Prior works on IRS-aided MIMO DFRC secrecy assume a single ED/target with either unicast or broadcast transmissions. Further, the indirect path of target backscatter via IRS is often ignored in these studies \cite{Deligiannis_Secrecy_Radar_TAES18,fangsinr}. In this paper, we generalize the DFRC interception problem to \textit{opt}imizing \textit{m}ultiple ED \textit{m}ulticast \textit{m}ulti-antenna \textit{sec}recy rate (OptM3Sec). Our physical layer design approach maximizes the secrecy rate while meeting power constraints and  maintaining a certain SINR for each radar target. In particular, we control the SINR at the radar target (ED SINR) by embedding in the transmit waveform a noise-like signal. The signal and the noise are precoded separately at the radar. We consider both direct and indirect (via IRS) paths for line-of-sight (LoS) and non-line-of-sight (NLoS) targets. We solve the resulting optimization problem with respect to the precoding matrices following the strategy adopted from \cite{fangsinr}.

\vspace{-12pt}
\section{System Model}
\label{sec:sysmod}
\begin{figure}[t]
\centering
\includegraphics[width=0.75\columnwidth]{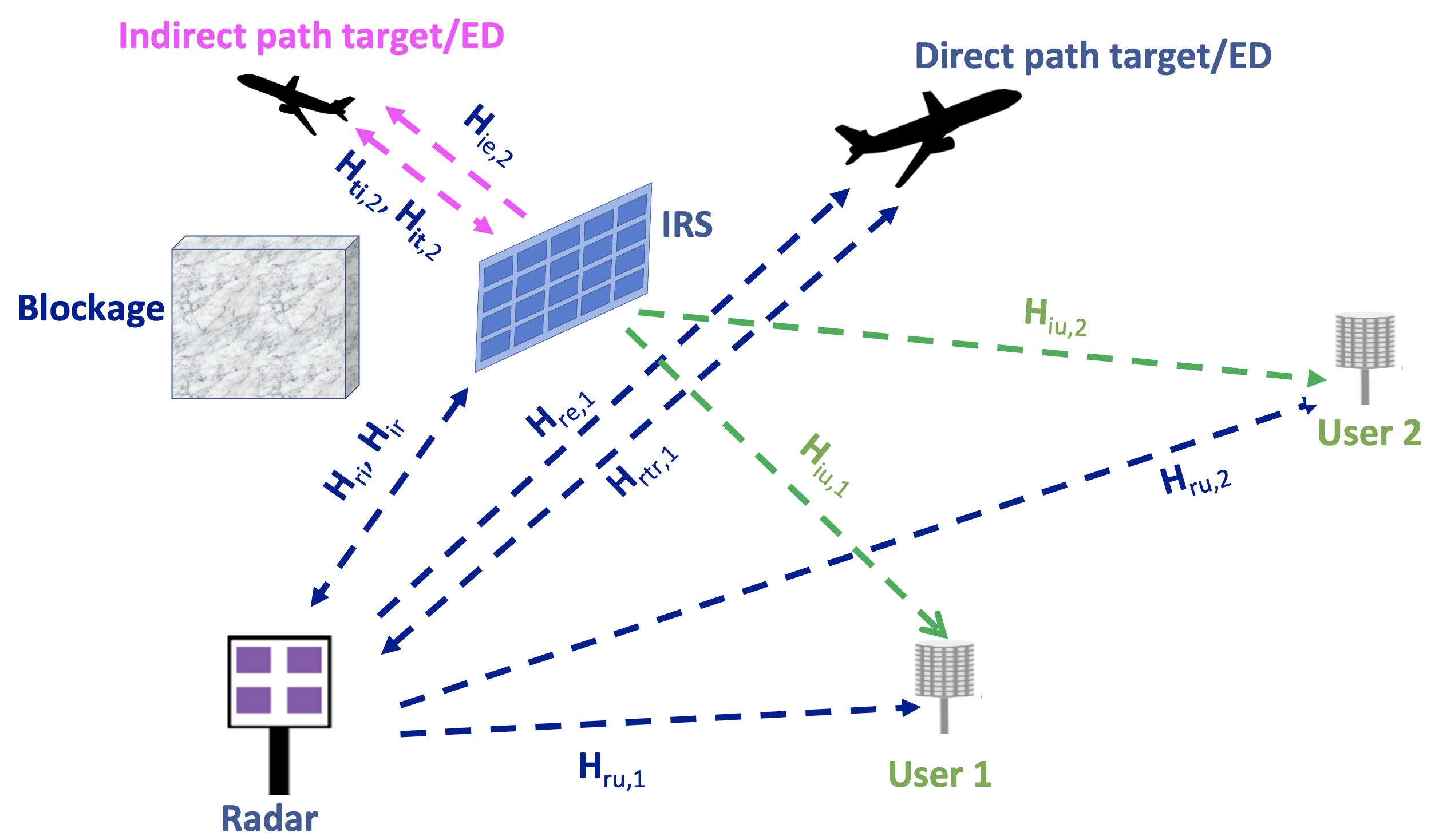}
\caption{MIMO-DFRC system with two targets/EDs and two users. \vspace{-12pt}}
\label{fig:sysmod}
\end{figure}

Consider a DFRC system comprising a MIMO radar, $L$ legitimate multi-antenna communications users, an IRS, and $K$ targets which are also EDs with multiple antennas (Fig.~\ref{fig:sysmod}). Assume that the MIMO radar has $N_T$ transmit and receive antennas, each legitimate user is equipped with $N_R$ receive antennas, and each ED receive antenna array has $N_E$ elements. The IRS is equipped with a $\sqrt{N} \times \sqrt{N}$ square array of reflecting units. 
A number of channels in this DFRC system are of our interest. Denote the channel gain matrix for the channel from radar to $k$-th target and back to the radar as
	\begin{equation}
		 \mathbf{H}_{rtr,k} = \beta_{k} \mathbf{a}_R(\theta _{k}) \mathbf{a}_T(\theta _{k})^{T} \in \mathbb{C}^{N_T \times N_T},
	\end{equation} 	
	where $\beta_{k}$ is the complex reflectivity that depends on the atmospheric attenuation and target's radar cross-section (RCS), $\theta _{k}$ is the azimuthal location of the target with respect to the radar, $\mathbf{a}_T(\theta _{k})$ ($ \mathbf{a}_R(\theta _{k})$) is the transmit (receive) steering vector of $k$-th target defined as
	\begin{eqnarray*} 
		\mathbf{a}_T(\theta _{k}) &=& [1, e^{j\frac{2\pi }{\lambda } d_r \text{sin}(\theta _{k})}, \ldots, e^{j\frac{2\pi }{\lambda } (N_T-1) d_r \text{sin}(\theta _{k})}]^T \\ \mathbf{a}_R(\theta _{k}) &=& [1, e^{j\frac{2\pi }{\lambda } d_r \text{sin}(\theta _{k})}, \ldots, e^{j\frac{2\pi }{\lambda } (N_T-1) d_r \text{sin}(\theta _{k})}]^T,
	\end{eqnarray*} 
where $d_r$ is the spacing between the antenna elements at the radar and $\lambda$ is the transmit signal wavelength. Similarly, denote the other channel matrices as $\bm{H}_{ri} \in \mathbb{C}^{N \times N}$ for radar-IRS; $\bm{H}_{ru,l} \in \mathbb{C}^{  N_R \times N_T}$ for radar-user; $\bm{H}_{re,k} \in \mathbb{C}^{ N_E \times N_T}$ for radar-ED; $\bm{H}_{iu,l} \in \mathbb{C}^{  N_R \times N}$ for IRS-user; $\bm{H}_{it,k} \in \mathbb{C}^{  1 \times N}$ for IRS-target; $\bm{H}_{ie,k} \in \mathbb{C}^{ N_E \times N}$ for IRS-ED; $\bm{H}_{ir} \in \mathbb{C}^{ N_T \times N}$ for IRS-radar; and $\bm{H}_{ti,k} \in \mathbb{C}^{  N \times 1}$ for target-IRS paths.

The radar transmits information bearing signal $\mathbf{m}(t) \in \mathbb{C}^{K \times 1}$ and artificial noise (AN)  $\mathbf{s}(t) \sim \mathcal{CN}(\bm{0},\bm{I}) \in \mathbb{C}^{K \times 1}$ for jointly detecting the target and communicating with users. The AN is added to prevent targets eavesdropping on the information transmitted to the users. The transmit signal $\mathbf{x}(t)\in \mathbb{C}^{N_T \times 1}$ from the radar is
	\begin{equation}
	\mathbf{x}(t)=\mathbf{W} \mathbf{m}(t)+\mathbf{B} \mathbf{s}(t),
	\end{equation} 	
where $\mathbf {W} = [\mathbf{w}_{1},\mathbf{w}_{2},\cdots,\mathbf{w}_{K}] \in \mathbb{C}^{N_T \times K} $ is the precoding matrix for information and $\mathbf {B} = [\mathbf{b}_{1},\mathbf{b}_{2},\cdots,\mathbf{b}_{K}]\in \mathbb{C}^{N_T \times K}$ is the precoding matrix for the artificial noise.


Assume $\tau_{(\cdot),k}$ and $\omega_{(\cdot),k}$ are the range-time delay and Doppler shift corresponding to the target $k$ for a given channel (denoted by the first subscript), the continuous-time received signal at the radar is:
\begin{align}
&\mathbf{r}(t) = \sum_{k=1}^{K} \mathbf{H}_{rtr,k} \mathbf{x}(t-\tau_{rtr,k}) e^{j \omega_{rtr,k} t} \nonumber\\
& + \sum_{k=1}^{K}  \mathbf{H}_{tr,k} \mathbf{H}_{it,k} \Phi \mathbf{H}_{ri}  \mathbf{x}(t-\tau_{ritr,k}) e^{j \omega_{ritr,k} t} \nonumber\\
& + \sum_{k=1}^{K}  \mathbf{H}_{ir} \Phi \mathbf{H}_{ti,k}  \mathbf{H}_{rt,k} \mathbf{x}(t-\tau_{rtir,k}) e^{j \omega_{rtir,k} t} \nonumber\\
& + \sum_{k=1}^{K}  \mathbf{H}_{ir} \Phi \mathbf{H}_{ti,k} \mathbf{H}_{it,k} \Phi \mathbf{H}_{ri,k} \mathbf{x}(t-\tau_{ritir,k}) e^{j \omega_{ritir,k} t} +\bm{n}_T(t),
\end{align}
where $\mathbf{n}_T\sim \mathcal{CN}(\mathbf{0},\sigma_T^2 \mathbf{I})$ is the additive white Gaussian noise. The received signal from the radar-IRS-target-IRS-radar path is ignored here because it is much weaker and delayed than the other returns.

The received signal at the $k$-th ED is:
\begin{flalign}
\mathbf{z}_k(t) &=   \mathbf{H}_{re} (\mathbf{W} \mathbf{m}(t-\tau_{re,k})+\mathbf{B} \mathbf{s}(t-\tau_{re,k})) e^{j \omega_{re,k} t} \nonumber\\
& +  \mathbf{H}_{ie,k} \Phi  \mathbf{H}_{ri} (\mathbf{W} \mathbf{m}(t-\tau_{rie,k})+\mathbf{B} \mathbf{s}(t-\tau_{rie,k})) e^{j \omega_{rie,k} t} \nonumber\\
&+\bm{n}_{E,k}(t),
\end{flalign}	
where   $\mathbf{n}_{E,k}(t)\sim \mathcal{CN}(\mathbf{0},\sigma_{E,k}^2 \mathbf{I})$ is the noise at the $k$-th ED.

The received signal at the $l$-th user is:
\begin{flalign}
\mathbf{y}_l(t) 
&=   \mathbf{H}_{ru} (\mathbf{W} \mathbf{m}(t-\tau_{ru,l})+\mathbf{B} \mathbf{s}(t-\tau_{ru,l})) e^{j \omega_{ru,l} t} \nonumber\\
& +  \mathbf{H}_{iu,l} \Phi  \mathbf{H}_{ri} (\mathbf{W} \mathbf{m}(t-\tau_{riu,l})+\mathbf{B} \mathbf{s}(t-\tau_{riu,l})) e^{j \omega_{riu,l} t} \nonumber\\
&+\bm{n}_U(t),
\end{flalign}	
where   $\mathbf{n}_{U,k}\sim \mathcal{CN}(\mathbf{0},\sigma_{U,k}^2 \mathbf{I})$ is the noise at the $k$-th ED.

The channel matrix for the path between radar and $k$-th ED is
\begin{eqnarray*}
	\mathbf{G}_{k} = \alpha_{k} \mathbf{a}_{E,k}(\phi_{k}) \mathbf{a}_T(\theta _{k})^{T},
\end{eqnarray*} 
where	$\alpha_{k}$ is the path attenuation, $\phi_{k}$ is the direction of the radar from ED's reference, and $ \mathbf{a}_{E,k}(\phi_{k})$ is the receive steering vector at target $k$ defined as
\begin{eqnarray*} 
	\mathbf{a}_{E,k}(\phi_{k}) &=& [1, e^{j\frac{2\pi }{\lambda } d_e \text{sin}(\phi _{k})}, \ldots, e^{j\frac{2\pi }{\lambda } (N_E-1) d_e \text{sin}(\phi_{k})}]^T 
\end{eqnarray*} 
The channel matrix between radar and $i$-th legitimate receiver is $\mathbf{F}_{i}$.

Our goal is to determine the optimal precoding matrices $\mathbf{W} $  and $\mathbf{B}$ that maximize the secrecy rate and target detection \cite{Deligiannis_Secrecy_Radar_TAES18}. Typical physical layer methods unrealistically assume that the channel to the ED is known. However, in the DFRC scenario, one could detect all the targets and estimate their channels. Thus, all but the legitimate receiver are classified as EDs with known channels.
\vspace{-12pt}
\section{Multicast with Multiple EDs}
\label{sec:multi}
Assume that the received signal at the radar $\mathbf{r}(t)$ is compensated for time delays and Doppler shifts and processed with matched filters corresponding to different targets \cite{Friedlander_Beamforming_Radar_TAES12}. 
The targets are sufficiently far apart so as not to interfere with each other's responses. We partition the targets into two sets $\mathcal{D}$ and $\mathcal{I}$, such that $|\mathcal{D}|+|\mathcal{I}| = K$, for which only direct and indirect paths are available, respectively.

The signal returning via the IRS is delayed and very weak. Therefore, it can be neglected if the direct path is available. Denote $\mathbf{H}_{dc,k} = \mathbf{H}_{rtr,k}$ as the direct path channel for the $k$-th target, $k\in\mathcal{D}$. Then, the SINR corresponding to the $k$-th target with beamformer weights $\mathbf {w}_{k}$ and $\mathbf {b}_{k}$ is
\begin{align} \text{SINR}_{dc,k} &= \frac{\textrm{Tr}( \mathbf{H}_{dc,k} \mathbf{w}_{k} \mathbf{w}_{k}^H \mathbf{H}_{dc,k}^{H}) + \textrm{Tr}(\mathbf{H}_{dc,k} \mathbf{b}_{k} \mathbf{b}_{k}^H \mathbf{H}_{dc,k}^{H}) }{ \sigma _{T}^2},\nonumber\\
&\;\;\;\;\;\;\;\;\;\;\;\;\;\;\;\;\;\;\;\;\;\;\;\;\;\;\;\;\;\;\;\;\;\;\;\;\;\;\;\;\;\;\;\;\;\;\;\;\;\;\;\;\;\;\;\;\;\;\;\; k\in\mathcal{D}.
\end{align}
However, if there is an obstacle in the direct path, then the direct path is ignored and we consider the following definition for targets $k\in\mathcal{I}$:
\begin{align}
    \text{SINR}_{in,k} &= \frac{\textrm{Tr}(\mathbf {H}_{in,k} \mathbf {w}_{k} \mathbf {w}_{k}^H \mathbf {H}_{in,k}^{H}) + \textrm{Tr}(\mathbf {H}_{in,k} \mathbf{b}_{k} \mathbf {b}_{k}^H \mathbf {H}_{in,k}^{H}) }{ \sigma _{T}^2},\nonumber\\
&\;\;\;\;\;\;\;\;\;\;\;\;\;\;\;\;\;\;\;\;\;\;\;\;\;\;\;\;\;\;\;\;\;\;\;\;\;\;\;\;\;\;\;\;\;\;\;\;\;\;\;\;\;\;\;\;\;\;\;\;k\in\mathcal{I},
\end{align}
where $\mathbf{H}_{in,k} = \mathbf{H}_{ir} \boldsymbol{\Phi} \mathbf{H}_{ti,k} \mathbf{H}_{it,k} \boldsymbol{\Phi} \mathbf{H}_{ri,k}$ is the channel gain for the indirect path for the $k$-th target.


From received signal $\mathbf{z}_k(t)$ at the $k$-th ED, its rate is
\footnotesize
\begin{equation}
	R_{e,k}=\log \det \left( \mathbf{I} + \left(\sigma _{E,k}^2 \mathbf{I} + \mathbf{G}_{k}  \mathbf{B} \mathbf{B}^H \mathbf{G}_{k}^{H}  \right)^{-1} \left(\mathbf{G}_{k} \mathbf {W} \mathbf {W}^H \mathbf{G}_{k}^{H}\right) \right). 
\end{equation} 
\normalsize

\noindent In the multicast setting, the radar transmits a common message $m$ to all the users. Let $\mathbf{F}_{l}$ denote the channel matrix between the radar and user $l$, $m_l(t)$ denote the message for the $l$-th user, $l=1,2,\cdots,L$ and $\mathbf{m}(t)= [m_1(t),m_2(t),\cdots,m_K(t)]^T $. The received signal at the $l$-th user is 
	\begin{equation}
		\mathbf{y}_l(t)=\mathbf{F}_i(\mathbf{W} \mathbf{m}(t)+\mathbf{B} \mathbf{s}(t)) + \mathbf{n}_l(t)
	\end{equation}
where $\mathbf{n}_l\sim \mathcal{CN}(\mathbf{0},\sigma_l^2 \mathbf{I})$ is the receiver noise. Hence, the rate at the $l$-th user is \cite{Liang_IT_Security_FnT09} 
\begin{equation}
	R_{u,l}=\log \det \left( \mathbf{I} + \left(\sigma _{R,l}^2 \mathbf{I} + \mathbf{F}_l  \mathbf{B} \mathbf{B}^H \mathbf {F}_i^{H}  \right)^{-1} \left(\mathbf {F}_l \mathbf {W} \mathbf {W}^H \mathbf {F}_l^{H}\right) \right). 
\end{equation} 
The secrecy rate is maximized by solving the optimization problem:
\begin{align}
		&\underset{\mathbf{W},\mathbf{B}}{\textrm{{maximize}}} \;
		 \underset{l,k}{\textrm{minimize}} \; [R_{u,l} - R_{e,k}]^+  \nonumber\\
		& \text{subject to}\;\;\; 
		   \text{Tr}(\mathbf{B} \mathbf{B}^H) + \text{Tr}(\mathbf{W} \mathbf{W}^H) \leq P,\nonumber\\
		& 	\textrm{Tr}(\mathbf{H}_{dc,k} \mathbf{w}_{k} \mathbf{w}_{k}^H \mathbf{H}_{dc,k}^{H}) + \textrm{Tr}( \mathbf{H}_{dc,k} \mathbf{b}_{k} \mathbf {b}_{k}^H \mathbf{H}_{dc,k}^{H}) \geq \gamma_k,  k\in \mathcal{D},\nonumber\\
		& 	\textrm{Tr}( \mathbf{H}_{in,k} \mathbf{w}_{k} \mathbf{w}_{k}^H \mathbf{H}_{in,k}^{H}) + \textrm{Tr}( \mathbf{H}_{in,k} \mathbf{b}_{k} \mathbf {b}_{k}^H \mathbf{H}_{in,k}^{H}) \geq \gamma_k, k\in \mathcal{I}, \nonumber\\
		\label{eqn:SR_MU_Broadcast}
\end{align}
where $P$ denotes the transmit power constraint and $\gamma_k$ denotes the SINR threshold for target $k$.

\vspace{-12pt}
\section{Secrecy Rate Maximization}
\label{sec:SR}
In order to solve \eqref{eqn:modified-optimization}, our OptM3Sec algorithm follows a three-step approach as in \cite{fangsinr}. However, unlike \cite{fangsinr}, our algorithm updates $\mathbf{W}$ and $\mathbf{B}$ through a nonconvex optimization problem, whose nonconvex feasible region is convexified via linearization of the SNR constraints. Further,the optimization over $\boldsymbol{\Phi}$ is highly nonconvex in our problem and, therefore we use a specific version of stochastic gradient ascent as opposed to the semidefinite relaxation in \cite{fangsinr}. Our algorithm consists of the following steps. 
\\
\\
\noindent {\bf Optimizing over auxiliary matrices:} 
Rewrite \eqref{eqn:SR_MU_Broadcast} as 
 \begin{align}\label{eqn:modified-optimization}
		&\underset{\mathbf{W},\mathbf{B}}{\textrm{{maximize}}} \; \lambda \nonumber\\
		& \text{subject to}\;\;\; 
		   \text{Tr}(\mathbf{B} \mathbf{B}^H) + \text{Tr}(\mathbf{W} \mathbf{W}^H) \leq P,\nonumber\\
		& 	\textrm{Tr}( \mathbf{H}_{dc,k} \mathbf{w}_{k} \mathbf{w}_{k}^H \mathbf{H}_{dc,k}^{H}) + \textrm{Tr}( \mathbf{H}_{dc,k} \mathbf{b}_{k} \mathbf {b}_{k}^H \mathbf{H}_{dc,k}^{H}) \geq \gamma_k,  k\in \mathcal{D},\nonumber\\
		& 	\textrm{Tr}( \mathbf{H}_{in,k} \mathbf{w}_{k} \mathbf{w}_{k}^H \mathbf{H}_{in,k}^{H}) + \textrm{Tr}( \mathbf{H}_{in,k} \mathbf{b}_{k} \mathbf {b}_{k}^H \mathbf{H}_{in,k}^{H}) \geq \gamma_k, k\in \mathcal{I},\nonumber\\
		& R_{u,l} - R_{e,k} \geq \lambda \geq 0 \, \forall l,k.
\end{align}

After some tedious algebra, we obtain
\begin{flalign}
&R_{u,l} - R_{e,k} \nonumber\\
&=\underbrace{\log \det \left( \mathbf{I} + \mathbf {F}_l \mathbf {W} \mathbf {W}^H \mathbf {F}_l^{H} \left(\sigma _{R,l}^2 \mathbf{I} + \mathbf{F}_l  \mathbf{B} \mathbf{B}^H \mathbf {F}_i^{H}  \right)^{-1}  \right)}_{\doteq G_1} \nonumber\\
& + \underbrace{\log \det   \left(\sigma _{E,k}^2 \mathbf{I} + \mathbf{G}_{k}  \mathbf{B} \mathbf{B}^H \mathbf{G}_{k}^{H}  \right)}_{\doteq G_2} \nonumber\\
&- \underbrace{\log \det \left(  \sigma _{E,k}^2 \mathbf{I} + \mathbf{G}_{k}  \mathbf{B} \mathbf{B}^H \mathbf{G}_{k}^{H}   + \mathbf{G}_{k} \mathbf {W} \mathbf {W}^H \mathbf{G}_{k}^{H} \right)}_{\doteq G_3}. \label{eqn:definition-of-G}
\end{flalign}
It follows from \cite[Lemma~$4.1$]{shi2015secure} that
\begin{align*}
G_1 &=  \underset{\mathbf{W}_b \succ \bm{0, }\mathbf{U}_b}{\mathrm{max}} \left( \log \det (\mathbf{W}_b)- \textrm{Tr}(\mathbf{W}_b \mathbf{E}_b(\mathbf{U}_b, \mathbf{W}, \mathbf{B})) \right)\nonumber\\
&\;\;\;\;\;\;\;\;+\mathrm{constant},
\end{align*}
where  $\bm{W}_b, \bm{U}_b$ are auxialiary matrices of appropriate dimensions, and 
\begin{eqnarray*}
 \mathbf{E}_b(\mathbf{U}_b, \mathbf{W}, \mathbf{B}) & \doteq & (\bm{I}-\bm{U}_b^H \bm{F}_l \bm{W}) (\bm{I}-\bm{U}_b^H \bm{F}_l \bm{W})^H \\
&& + \bm{U}_b^H \left(\sigma _{R,l}^2 \mathbf{I} + \mathbf{F}_l  \mathbf{B} \mathbf{B}^H \mathbf {F}_i^{H}  \right) \bm{U}_b.
\end{eqnarray*}

The optimal solution $(\bm{W}_{b,l,k}^*, \bm{U}_{b,l,k}^*)$ is
\begin{align}\label{eqn:optimal-U-for-G1}
\bm{U}_{b,l,k}^* &= \underset{\mathbf{U}_b}{\arg \max} \left( \log \det (\mathbf{W}_b)- \textrm{Tr}(\mathbf{W}_b \mathbf{E}_b(\mathbf{U}_b, \mathbf{W}, \mathbf{B})) \right) \nonumber\\
&= (  \sigma _{R,l}^2 \mathbf{I} + \mathbf{F}_l  \mathbf{B} \mathbf{B}^H \mathbf {F}_l^{H} +  \bm{F}_l \bm{W} \bm{W}^H  \bm{F}_l^H  )^{-1}  \bm{F}_l \bm{W},
\end{align}
and 
\begin{equation}\label{eqn:optimal-W-for-G1}
\bm{W}_{b,l,k}^* = ( \mathbf{E}_b(\mathbf{U}_{b,l,k}^*, \mathbf{W}, \mathbf{B}) )^{-1}.
\end{equation}
Similarly, we use auxiliary matrices $\bm{W}_e, \bm{U}_e$ to define
\begin{equation*}
G_2=  \underset{\mathbf{W}_e \succ \bm{0, }\mathbf{U}_e}{\mathrm{max}} \left( \log \det (\mathbf{W}_e)- \textrm{Tr}(\mathbf{W}_e \mathbf{E}_e(\mathbf{U}_e, \mathbf{B})) \right)+\mathrm{constant},
\end{equation*}
where 
\begin{align}
 \mathbf{E}_e(\mathbf{U}_e,   \mathbf{B}) &\doteq  \left(\bm{I}-\frac{1}{\sigma_{E,k}}\bm{U}_e^H \bm{G}_k \bm{B}\right) \left(\bm{I}-\frac{1}{\sigma_{E,k}} \bm{U}_e^H \bm{G}_k \bm{B}\right)^H\nonumber\\
 &\;\;\;\;\;\;\; + \bm{U}_e^H  \bm{U}_e.
\end{align}
\noindent This yields
\begin{eqnarray}\label{eqn:optimal-U-for-G2}
\bm{U}_{e,l,k}^* 
&=& \left( \mathbf{I} + \frac{1}{\sigma_{E,k}^2} \mathbf{G}_{k}  \mathbf{B} \mathbf{B}^H \mathbf{G}_{k}^{H}  \right)^{-1} \frac{\bm{G}_k}{\sigma_{E,k}}  \bm{B},
\end{eqnarray}
and 
\begin{equation}\label{eqn:optimal-W-for-G2}
\bm{W}_{e,l,k}^* = ( \mathbf{E}_e(\mathbf{U}_{e,l,k}^*,   \mathbf{B}) )^{-1}.
\end{equation}
Similarly, define
\begin{equation*}
G_3=  \underset{\mathbf{W}_z \succ \bm{0} }{\mathrm{max}} \left( \log \det (\mathbf{W}_z)- \textrm{Tr}(\mathbf{W}_z \mathbf{E}_z(\mathbf{W}, \mathbf{B})) \right)+\mathrm{constant},
\end{equation*}
 and $\bm{W}_{z,l,k}^*=\left(   \mathbf{I} + \frac{1}{\sigma_{E,k}^2}  \mathbf{G}_{k}  \mathbf{B} \mathbf{B}^H \mathbf{G}_{k}^{H}   + \frac{1}{\sigma_{E,k}^2} \mathbf{G}_{k} \mathbf {W} \mathbf {W}^H \mathbf{G}_{k}^{H} \right)^{-1}.$
 \noindent Define $\bm{T} \doteq \{\bm{U}_b, \bm{W}_b, \bm{U}_e, \bm{W}_e, \bm{W}_z, \bm{W}, \bm{B}, \bm{\Phi}\}$ as the collection of all unknown parameters. We solve \eqref{eqn:modified-optimization} using the block coordinate descent (BCD) method.
 \\\\
\noindent {\bf Optimizing over $(\mathbf{W}, \mathbf{B})$ given $\boldsymbol{\Phi}$: } 
 Here, the SNR and secrecy rate constraints in \eqref{eqn:modified-optimization} lead to a non-convex feasible region. 
The secrecy rate constraint can be convexified by \eqref{eqn:secrecy-rate-in-terms-of-beamforming-matrices}. 
 \begin{figure*}[t!]
  \footnotesize
\begin{eqnarray}\label{eqn:secrecy-rate-in-terms-of-beamforming-matrices}
&& R_{u,l} - R_{e,k} \nonumber\\
&=& \left( \log \det (\mathbf{W}_b)- \textrm{Tr}(\mathbf{W}_b \mathbf{E}_b(\mathbf{U}_b, \mathbf{W}, \mathbf{B})) \right)  +    \left( \log \det (\mathbf{W}_e)- \textrm{Tr}(\mathbf{W}_e \mathbf{E}_e(\mathbf{U}_e, \mathbf{B})) \right)  +    \left( \log \det (\mathbf{W}_z)- \textrm{Tr}(\mathbf{W}_z \mathbf{E}_z(\mathbf{W}, \mathbf{B})) \right) + constant \nonumber\\
&=& Tr \left( \bm{W}_b \left( \bm{U}_b^H \bm{F}_l \bm{W} + \bm{W}^H \bm{F}_l^H \bm{U}_b \right) \right) -Tr \left( \bm{W}_b \bm{U}_b^H \bm{F}_l \bm{W} \bm{W}^H \bm{F}_l^H \bm{U}_b \right) - \frac{1}{\sigma _{E,k}^2} Tr \left( \bm{W}_z \mathbf{G}_{k} \mathbf {W} \mathbf {W}^H \mathbf{G}_{k}^{H}  \right) \nonumber\\
&& + Tr \left(\bm{W}_e \left( \bm{U}_e^H \bm{G}_k \bm{B} + \bm{B}^H \bm{G}_k^H \bm{U}_e \right) \right) - Tr \left( \bm{W}_b \bm{U}_b^H \bm{F}_l \bm{B} \bm{B}^H \bm{F}_l^H \bm{U}_b \right)- Tr \left(\bm{W}_e \bm{U}_e^H \bm{G}_k \bm{B} \bm{B}^H \bm{G}_k^H \bm{U}_e\right) - Tr \left( \frac{1}{\sigma _{E,k}^2} \bm{W}_z \mathbf{G}_{k}  \mathbf{B} \mathbf{B}^H \mathbf{G}_{k}^{H} \right) \nonumber\\
&&+ g(\bm{W}_b, \bm{W}_e, \bm{W}_z)+\mathrm{constant}.
\end{eqnarray}\hrule
\normalsize
\end{figure*}
We linearize the SNR constraints  by first-order Taylor series approximation around some initial approximations $\mathbf{\tilde{w}}_k$ and $\mathbf{\tilde{b}}_k$. Define
$f_{dc}(\mathbf{w}_{k}, \mathbf{b}_{k} ) = \text{Tr} (\mathbf{H}_{dc,k} \mathbf{w}_{k} \mathbf{w}_{k}^H \mathbf{H}_{dc,k}^{H}) + \textrm{Tr}( \mathbf{H}_{dc,k} \mathbf{b}_{k} \mathbf {b}_{k}^H \mathbf{H}_{dc,k}^{H}) $. Then,
$$f_{dc}(\mathbf{w}_{k}, \mathbf{b}_{k} ) \approx f_{dc}(\mathbf{\tilde{w}_{k}}, \mathbf{\tilde{b}_{k} })  + \textrm{Re}(\nabla^T_{\mathbf{w}_{k}, \mathbf{b}_{k}} f_{dc}(\mathbf{\tilde{w}}_k, \mathbf{\tilde{b} }_k) (\mathbf{d}_k - \mathbf{\tilde{d}}_k)),$$
where $\mathbf{d}_k = [\mathbf{w}_k^T\;\mathbf{b}_k^T]^T$,
$\nabla_{\mathbf{w}_k, \mathbf{b}_k}f_{dc} = [\frac{ \partial f_{dc}^T}{\partial \mathbf{w}_k^H} \;\; \frac{ \partial f_{dc}^T}{\partial \mathbf{b}_k^H}]^T$, and $\frac{ \partial f_{dc}}{\partial \mathbf{w}_k^H} = 2 \mathbf{w}_{k}^H\mathbf{H}_{dc,k} \mathbf{H}_{dc,k}^H  $, and $\frac{ \partial f_{dc}}{\partial \mathbf{b}_k^H} = 2 \mathbf{b}_{k}^H\mathbf{H}_{dc,k} \mathbf{H}_{dc,k}^H  $. This yields

 \begin{align}
		&\underset{\mathbf{W},\mathbf{B}}{\textrm{{maximize}}} \; \lambda \nonumber\\
		& \text{subject to}\;\;\; 
		   \text{Tr}(\mathbf{B} \mathbf{B}^H) + \text{Tr}(\mathbf{W} \mathbf{W}^H) \leq P,\nonumber\\
		&  f_{dc}(\mathbf{\tilde{w}_{k}}, \mathbf{\tilde{b}_{k} })  + \textrm{Re}(\nabla^T_{\mathbf{w}_{k}, \mathbf{b}_{k}} f_{dc}(\mathbf{\tilde{w}}_k, \mathbf{\tilde{b} }_k) (\mathbf{d}_k - \mathbf{\tilde{d}}_k)) \geq \gamma_k,  k\in \mathcal{D},\nonumber\\
		&  f_{in}(\mathbf{\tilde{w}_{k}}, \mathbf{\tilde{b}_{k} })  + \textrm{Re}(\nabla^T_{\mathbf{w}_{k}, \mathbf{b}_{k}} f_{in}(\mathbf{\tilde{w}}_k, \mathbf{\tilde{b} }_k) (\mathbf{d}_k - \mathbf{\tilde{d}}_k)) \geq \gamma_k, k\in \mathcal{I},\nonumber\\
		& R_{u,l} - R_{e,k} \geq \lambda \geq 0 \, \forall l,k.
	\label{eqn:SR_MU_linearised-optimization-over-W-B}
\end{align}
\noindent This convex optimization problem is solved by any standard solver.
\\\\
\noindent {\bf Optimizing $\bm{\Phi}$ for given $\bm{W}, \bm{B}$: }
The SNR constraint for the direct radar-target-radar links and the secrecy rates in \eqref{eqn:modified-optimization} do not depend on $\boldsymbol{\Phi}$. Denote $h(\boldsymbol{\Phi}) = \mathbf{H}_{in,k} \mathbf{w}_{k} \mathbf{w}_{k}^H   \mathbf{H}_{in,k}^{H} +  \mathbf{H}_{in,k} \mathbf{b}_{k} \mathbf {b}_{k}^H \mathbf{H}_{in,k}^{H}$. This results in the following optimization problem
\begin{equation}\label{eqn:phi-optimization}
    \underset{\boldsymbol{\Phi}}{\textrm{maximize}} \,\,\,  \underbrace{ \underset{k \in \mathcal{I}}{\textrm{minimize}}\; \textrm{Tr}( h(\boldsymbol{\Phi})) - \gamma_k }_{\doteq f(\boldsymbol{\Phi}) = \tilde{f}(\tilde{\boldsymbol{\Phi}})}.
\end{equation}

\noindent Denote the phase shift induced by the $i$-th reflection unit of the IRS as $\phi_i \in [0,\pi]$ and $\tilde{\boldsymbol{\Phi}} \doteq [\phi_1 \,\,  \phi_2 \,\, \cdots  \phi_N]^T$. Since the objective function $\tilde{f}(\tilde{\boldsymbol{\Phi}})$ is nonconvex in $(\tilde{\boldsymbol{\Phi}})$, we solve it by simultaneous perturbation stochastic approximation (SPSA) \cite{spall1992multivariate}, which is a specific version of stochastic gradient ascent. In SPSA, we iteratively update $\tilde{\boldsymbol{\Phi}}(t)$. In the $t$-th iteration, a zero-mean perturbation vector $\boldsymbol{\Delta}(t) \in \mathbb{R}^{N \times 1}$ is generated independently, where each of its entries $\{\Delta_i(t): 1 \leq i \leq N\}$ is chosen from the set $\{-1,1\}$ with equal probability. Then the iterate $\tilde{\boldsymbol{\Phi}}(t)$ is perturbed in two opposite directions as $\tilde{\boldsymbol{\Phi}}^+(t) \doteq \tilde{\boldsymbol{\Phi}}(t)+ c(t) \boldsymbol{\Delta}(t)$ and $\tilde{\boldsymbol{\Phi}}^-(t) \doteq \tilde{\boldsymbol{\Phi}}(t)- c(t) \boldsymbol{\Delta}(t)$. Next, for all $1 \leq i \leq N$, the $i$-th component of $\tilde{\boldsymbol{\Phi}}(t)$ is updated as
\begin{equation}
    \phi_i(t+1)=\phi_i(t)+ a(t) \times \frac{\tilde{f}(\tilde{\boldsymbol{\Phi}}^+(t))-\tilde{f}(\tilde{\boldsymbol{\Phi}}^-(t))}{2 c(t) \Delta_i(t)},
\end{equation}
and the iterates are projected onto the interval $[0,\pi]$ to ensure feasibility. The {\em positive} step size sequences $\{a(t)\}_{t \geq 1}$ and $\{c(t)\}_{t \geq 1}$ need to satisfy the following conditions: (i) $\sum_{t=1}^{\infty}a(t)=\infty$, (ii) $\sum_{t=1}^{\infty}a^2(t)<\infty$, (iii) $\underset{t \rightarrow \infty}{\lim} c(t)=0$, and (iv) $\sum_{t=1}^{\infty} \frac{a^2(t)}{c^2(t)} < \infty$. The SPSA iteration is run until a suitable stopping criterion is met. Algorithm~\ref{algorithm:OPTM3SEC} summarizes the steps of our OptM3Sec method.

\begin{algorithm}[H]
	\caption{\textit{Opt}imization of \textit{m}ultiple ED \textit{m}ulticast \textit{m}ulti-antenna DFRC \textit{sec}recy rate (OptM3Sec)}
	\label{algorithm:OPTM3SEC}
    \begin{algorithmic}[1]
    \Statex \textbf{Input:} All channel gains and noise covariances, $P$,  $\{\gamma_k: k \in \mathcal{D}, k \in \mathcal{I}\}$.
    \Statex \textbf{Output:} $\mathbf{W}$, $\mathbf{B}$, $\tilde{\boldsymbol{\Phi}}$
    \State {\bf Initialisation:} $\mathbf{W}(0), \mathbf{B}(0), \tilde{\boldsymbol{\Phi}}(0)$, and $\tau=0$.
    \For{$\tau = 1,2,3, \cdots$} \State  Given $\mathbf{W}(\tau-1)$, $\mathbf{B}(\tau-1)$, $\tilde{\boldsymbol{\Phi}}(\tau-1)$, compute $\bm{U}_{b,l,k}^*(\tau)$, $\bm{W}_{b,l,k}^*(\tau)$, $\bm{U}_{e,l,k}^*(\tau)$, $\bm{W}_{e,l,k}^*(\tau)$, $\bm{W}_{z,l,k}^*(\tau)$ for all $l,k$. 
        \State For given $\bm{U}_{b,l,k}^*(\tau)$, $\bm{W}_{b,l,k}^*(\tau)$, $\bm{U}_{e,l,k}^*(\tau)$, $\bm{W}_{e,l,k}^*(\tau)$, $\bm{W}_{z,l,k}^*(\tau)$ and $\tilde{\boldsymbol{\Phi}}(\tau-1)$, find $\mathbf{W}(\tau)$ and $\mathbf{B}(\tau)$ by solving \eqref{eqn:SR_MU_linearised-optimization-over-W-B}.
        \State  For given $\bm{U}_{b,l,k}^*(\tau)$, $\bm{W}_{b,l,k}^*(\tau)$, $\bm{U}_{e,l,k}^*(\tau)$, $\bm{W}_{e,l,k}^*(\tau)$, $\bm{W}_{z,l,k}^*(\tau)$ and given $\mathbf{W}^*(\tau)$ and $\mathbf{B}^*(\tau)$, find $\tilde{\boldsymbol{\Phi}}(\tau)$ by running many iterations of SPSA.
    \EndFor
    \State Stop when a suitable criterion is met.
	\end{algorithmic}
\end{algorithm}

\vspace{-12pt}
\section{Experiments and Summary}
\label{sec:numexp}
\begin{figure}[t]
\begin{centering}
\begin{center}
\includegraphics[width=0.75\columnwidth]{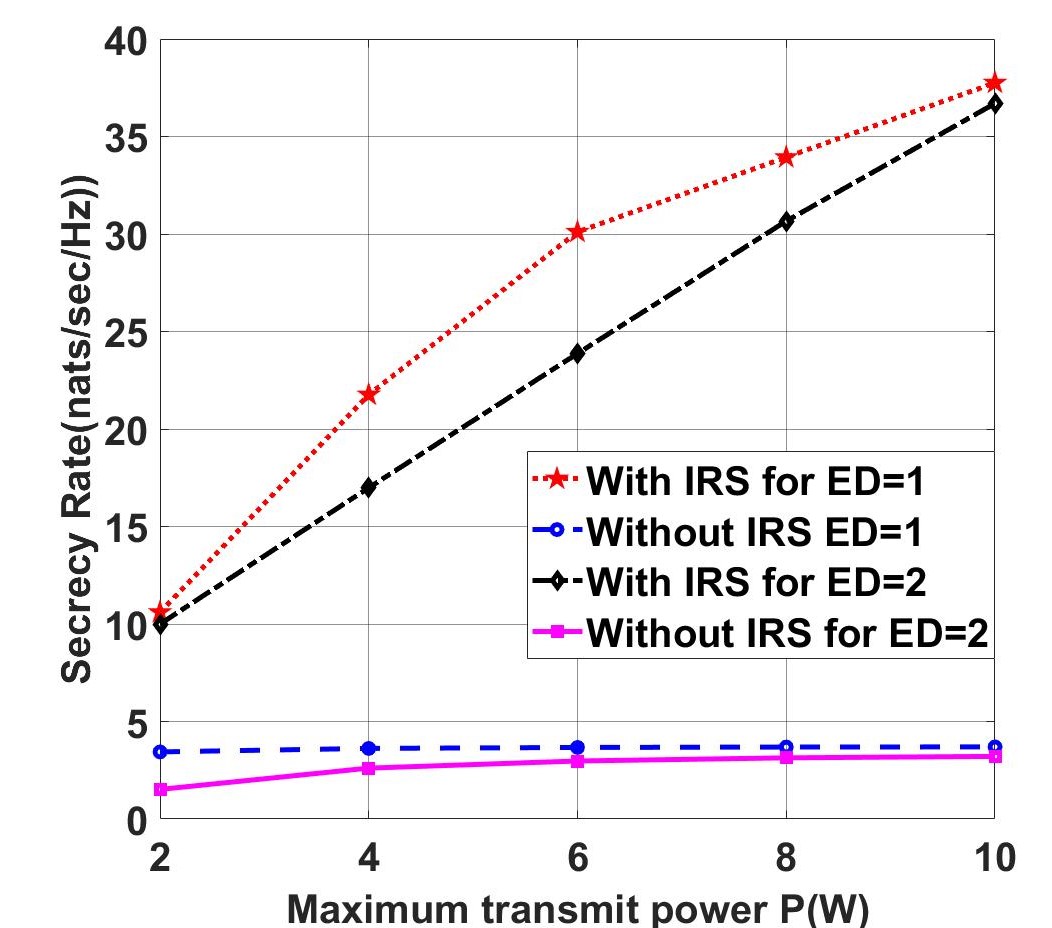}
\end{center}
\end{centering}
\vspace{-5mm}
\caption{Achieved Secrecy rate against different maximum power limits and for different ED/target.\vspace{-12pt} }
\label{fig:Single antenna model_ED}
\vspace{+2mm}
\end{figure}
We validated out proposed method through numerical experiments. We considered a system with a single IRS, a radar, two multiple antenna receivers. We set $N_T=4$ for radar transmit and receive antennas, $N_R=4$ for user receive antennas, and $N_E=4$ receive antennas for each ED. The IRS is equipped with $N=10$ array of reflecting units. We consider two attack scenarios for our simulations, the first being when the attack is performed by an single ED and the second when the attack is done by two EDs. In the first attack setup we consider the target's true directions with respect to the radar as $\theta_1=72^\circ$ and from the reference of two EDs, the radar is located at $\phi_1=-85^\circ$. In the second scenario the target's true directions with respect to the radar are $\theta_1=72^\circ$ and $\theta_2=78^\circ$. From the reference of two EDs, the radar is located at $\phi_1=-85^\circ$ and $\phi_2=-88^\circ$. The RCS coefficient and path loss variables are set as $\alpha_1=0.1$, $\alpha_2=0.1$, $\beta_1=0.1$, and $\beta_2=0.1$. The variances of all Gaussian noise variables are set to unity.

For all channels involved, we sampled each channel coefficient from an independent circularly symmetric complex Gaussian random variable with zero mean and variance of unity. Fig \ref{fig:Single antenna model_ED} shows that in presence of a single ED, with our OptM3Sec algorithm, the secrecy rate increases with the maximum transmit power $P$ compared to the case when there is no IRS present. It is also evident from Fig \ref{fig:Single antenna model_ED} that in presence of two EDs, the secrecy rate also increases with the maximum transmit power $P$, however it can be observed that due to the increase in the number of EDs there is a slight decrease in the secrecy rate which is very intuitive. The above results highlight the significance of IRS in secrecy rate performance for multicast IRS-aided MIMO DFRC system with multiple EDs. 

\bibliographystyle{IEEEtran}
\bibliography{main}
	
\end{document}